\documentstyle[12pt]{article}

\newcommand{\beq}{\begin{equation}}
\newcommand{\eeq}{\end{equation}}

\begin{document}

\rightline{{\bf  US--FT/6--96}}
\rightline{{\bf hep-ph/9602296}}

\begin{center}
\vskip 0.75cm

{\Huge {\bf Fusion of strings and cosmic rays}}\\
\vskip 0.2cm
{\Huge {\bf at ultrahigh energies}}\\
\vskip 0.8cm

N. Armesto, M. A. Braun$^{*}$, E. G. Ferreiro, C. Pajares and Yu. M. 
Shabelski$^{**}$ \\ 
{\sl Departamento de 
F\'{\i}sica de  Part\'{\i}culas,
Universidade de Santiago de
Compostela,\\ 15706--Santiago de Compostela, Spain}

\vskip 1.0truecm
{\Large {\bf Abstract}}
\end{center}
\begin{quotation}

It is shown that the fusion of strings is a source of particle production in
 nucleus--nucleus collisions outside the kinematical limits of nucleon--nucleon
collisions. This fact, together with another effect of string fusion, the 
reduction of multiplicities, sheds some light on two of the main problems 
of ultrahigh energy cosmic rays, the chemical composition and the energy of the
most energetic detected cosmic rays.

\end{quotation}
\vspace{3cm}
\begin{flushleft}
\end{flushleft}

\noindent $^{*}$ 
Permanent address: Department of High Energy
Physics, University of St. Petersburg, 198904 St. Petersburg, 
Russia.\\ 
\noindent $^{**}$ Permanent address: Petersburg Nuclear Physics Institute, 
Gatchina, 188350 St. Petersburg, Russia. 

\newpage

In the standard models of hadronic interactions 
(\cite{And87,Cap94,Wer93,Sor89}),
strings, chains or pomerons are exchanged between the projectile and 
target. 
The number of strings grows with the energy and with the number of 
nucleons 
of the participant nuclei. In the first approximation strings fragment 
into particles and resonances
in an independent way. However, the interaction between strings becomes 
important with their number
growing. This interaction has been introduced into some of the models 
(\cite{Paj90,And91,Sor92,Moh92,Aic93,Ran94}). In particular, fusion 
of strings has been
incorporated into the Dual Parton Model (DPM) (\cite{Moh92}) and 
the Quark Gluon String Model
 (QGSM) (\cite{Ame93}). Some of the effects of string
 fusion like  
strangeness and antibaryon
 enhancement (\cite{Arm94}), reduction of long range
 correlations
(\cite{Ele94}) 
and multiplicity suppression  have been studied comparing the results with the 
existing experimental data. Also  
predictions for the Relativistic Heavy Ion Collider (RHIC) and 
Large Hadron Collider 
(LHC) are avalaible. In this paper we explore another effect of 
string fusion, namely particle
production in nucleus--nucleus collisions outside the kinematical 
limits of nucleon--nucleon
collisions (the so--called cumulative effect).

It is shown that this effect is important already at present avalaible 
energies. A non negligible
number of baryons and mesons are produced with momenta greater than 
the ones of the
colliding nucleons. This effect,
together with the reduction of multiplicities, provides a natural 
explanation of some 
features of cosmic ray data like the rise of the average shower 
depth of maximum $X_{max}$  
(the amount of air penetrated
by the cascade when it reaches maximum size) (\cite{Bir93,Gai93}) 
with increasing energy  
from $10^{17}$ eV to 
$10^{19}$ eV, and the existence of events with energy above $10^{20}$ 
eV 
(\cite{Bir93,Efi90,Hay94}), higher than the expected
cut--off (\cite{Gre66,Ste68,Sig11}) due to the
scattering of cosmic rays with the microwave radiation background. 
Usually the first feature is explained by an enrichment of protons in 
the composition of
primary cosmic rays (\cite{Gai93}) as energy increases. However, 
as we show, if  
the composition of the primary cosmic rays is 
kept fixed in the energy range between 
$10^{17}$ eV and
$10^{19}$ eV, string fusion leads to a suppression of the multiplicity 
similar
to the one produced by changing heavy nuclei (Fe) by protons in the composition 
of the primary.
 On the other hand, since 
the momentum of a fused string comes from summing the momenta
 of its ancestor strings, 
 it is possible to obtain particles with more energy than the
initial nucleon--nucleon energy. Therefore, if
several strings fuse, the observed cosmic ray events 
with energy above
$10^{20}$ eV may actually correspond to three or four times less initial 
energy 
than that apparently measured.
 String fusion could make these events compatible with 
the existence of above mentioned cut--off.

To study these effects we use a Monte Carlo code based on the QGSM, 
in which the fusion of strings has
been incorporated (\cite{Ame93}). A detailed description of the Monte 
Carlo String Fusion Model (SFMC) and
comparison with experimental data can be found in Ref. 
\cite{Ame93,Arm94,Ele94}. 
 A hadron or nucleus collision is assumed to be an interaction between 
clouds of partons formed long
before the collision. Without string fusion partons are assumed to 
interact only once. Each
parton--parton interaction leads to the creation of colour strings. 
Since both the projectile and the
target must remain colourless, strings have to be formed in 
pairs. Hadrons and nuclei are
considered on the same footing. The nuclear wave function is taken as 
a convolution of the parton
distribution in a nucleon with the distribution of nucleons in the 
nucleus (given by the Wood--Saxon
shape).

 Strings fuse when their transverse positions come within
a certain interaction area, which is fixed previously to describe 
correctly the strangeness enhancement (\cite{Arm94}) shown by the data on
nucleus--nucleus collisions.
The fusion can take place only when the rapidity intervals of the strings
overlap.  It is formally described by allowing partons to interact several
times, the number of interactions being the same for projectile and 
target. The
quantum numbers of the fused string are determined by those of the interacting 
partons and
its energy--momentum is the sum of the energy--momentum of the ancestor 
strings.
The
colour charges of the fusing string ends sum into the  colour charge  
of the
resulting string ends according to the $SU(3)$ composition laws. 
In particular,
two triplet strings fuse into an antitriplet  and a sextet string, with
probabilities 1/3 and 2/3 respectively. A triplet and an antitriplet
string give rise to a singlet state and an octet string with 
probabilities 1/9
and 8/9 respectively. In present calculations only fusion of two strings
 is taken into account.

Particle production outside the nucleon--nucleon kinematical limits is 
a well known effect called
 cumulative effect, studied both theoretically and experimentally 
(\cite{Bal74,Str81,Efr88,Bra94,Bay79}). However at high enough 
energies, where the string picture can be applied, there are only data from 
one
collaboration at 400 GeV/c (\cite{Bay79}), with incoming protons against 
nuclei: Li, Be, C, 
Al, Cu and Ta. Generating p--A events in our code and comparing
the cumulative particle spectrum with these data we observe a 
reasonable agreement, as can be seen in Tables 1 and 2
where the invariant differential cross sections for production   
of protons, positive pions and positive kaons are shown.

 Passing to nucleus--nucleus collisions, 10000  
central S--S collisions and 1000 central Pb--Pb collisions 
were simulated at
 $\sqrt{s}=19.4$ AGeV.
 Also central Pb--Pb collisions at RHIC energies 
($\sqrt{s}=200$ AGeV) have been simulated. Distributions of baryons and
mesons in central S--S and Pb--Pb collisions at SPS energies with $x_F$ 
larger 
than 1 are  shown in Fig. 1 and 2.  
 The results for Pb--Pb collisions at $\sqrt{s}=200$ AGeV are very
 similar to the ones at $\sqrt{s}=19.4$ AGeV. In 1000 events  2015 
particles are found with  
$|x_F|$ larger than 1 to compare with 1783 at $\sqrt{s}=19.4$ AGeV. 
 This small change is due to a moderate increase of the number of 
strings with energy.

To study the case relevant for cosmic rays we simulated 1000
Fe--Air collisions at  $10^{17}$
eV (we used this energy and not $10^{20}$ eV to save computing 
time, rendering the simulation reliable). In this sample 198 particles with 
$|x_F|$ $>$ 1 were found. The average number of strings was found
to be 225, from
 which 62 joined to form  double strings. As mentioned, our code only
includes fusion of two strings. However we can estimate the number of 
strings participating in a triple fusion assuming that the probability
for triple fusion is roughly the square of that for double fusion. 
 Then one would expect that 18 strings join to form triple strings 
and 4 strings join to form a quadruple string. Therefore
the probability of obtaining particles with $|x_F|$ $>$ 2 or even 
$|x_F|$ $>$ 3 does not seem to be negligible. 
The energy around $3 \cdot
 10^{20}$
eV measured in several cosmic ray experiments could then be lowered by
a factor 2 to 4 if the described effect is
present and there are particles in the shower with $|x_F|$ $>$ 2 
or $|x_F|$ $>$ 3. This lower
energy
 for the primary may lie below  
the energy cut--off due to the scattering of cosmic
rays on the microwave background.

On the other hand, string fusion produces a suppression of 
multiplicities, which can
explain the rise of the average shower depth of maximum in cosmic rays 
as the energy increases, 
without requiring any change in the chemical composition.
It is usually accepted that there is a change in the cosmic ray chemical
composition between $10^{16}$ eV and $10^{19}$ eV. 
It seems that the composition becomes significantly lighter
with increasing energy, going from a heavy composition at $10^{16}$ eV to
a light one at energies higher than $10^{19}$ eV. 
According to this,
 the distribution of the shower depth of maximum
as a function of energy has been studied using a simple model 
of two components
(\cite{Gai93}), observing that
the change in the composition of the primary goes from approximatly  
 75 \% of iron component
and a 25 \% of proton
component at $10^{16}$ eV
 to  
 50 \% of iron
and a 50 \% of proton
at $10^{19}$ eV.
To study this point, we have computed the multiplicities of 
p--Air and Fe--Air interactions with and without string fusion in the
 whole range of energies studied (from $10^{16}$ to $10^{19}$ eV).
 As it can be seen in Fig. 3, with string fusion the 
multiplicity for a constant composition 
of 
 10 \% of proton and 90 \% of iron in the whole range of energy, 
essentially reproduces the multiplicity obtained 
without string fusion for a uniform 
change in the composition from 75 \% Fe and 25 \% proton at $10^{16}$ eV to 
50 \% Fe and 50 \% proton at $10^{19}$ eV. Thus the string fusion does the
same job as the composition change. 

Therefore, the change in the energy behaviour of the average 
shower depth of maximum $X_{max}$ can be due to a change in the 
interaction mechanism with
the existence of collective effects like string fusion, and not to a 
change in the chemical
composition of the primary cosmic rays. Further studies of this 
point would require combining the code used in this paper with the 
standard codes which describe the full cascade. 
 Work in this direction is in progress.

Summarizing, string fusion in nucleus--nucleus collisions produces a 
considerable number of 
particles outside the kinematical limits. This effect, together 
with the predicted suppression of multiplicities may help to understand 
the high energy cosmic rays
including the chemical composition of primaries and the bound on the highest
cosmic ray energy. Predictions of string fusion can be detected in future
experiments at RHIC, LHC and cosmic ray experiments (concretely the Auger 
proyect (\cite{Aug95})).

In conclusion we would like to thank N. S. Amelin, A. Capella, J. W. Cronin, 
G. Parente, J. Ranft and 
E. Zas for useful comments and discussions and the 
Comisi\'on Interministerial de Cienc\'{\i}a y Tecnolog\'{\i}a (CICYT)  
of Spain for financial support. M. A. Braun thanks
IBERDROLA, E. G. Ferreiro thanks
Xunta de Galicia and Yu. M. Shabelski 
the Direcci\'on General de Pol\'{\i}tica Cient\'{\i}fica of Spain 
for finantial support. 
This work was partially supported by the INTAS grant N 93--0079.

\newpage
\noindent{\Large {\bf Table captions}}

\vskip 0.5cm

\noindent{\bf Table 1.} 
Comparison of experimental data (\cite{Bay79}) 
on the invariant differential cross section 
${\sigma} = E \frac{d\sigma}{dp_3}$ for $\pi^{+}$ and $K^{+}$ vs $p$  
(GeV/c), laboratory angle $118^{o}$, $P_{lab}$
 = 400 GeV, for p--Li
and p--Ta collisions with  
the String Fusion Model code results, with and without string fusion.

\noindent{\bf Table 2.} 
Comparison of experimental data (\cite{Bay79}) 
on the invariant differential cross section
${\sigma} = E \frac{d\sigma}{dp_3}$ for protons vs $p$ 
(GeV/c), laboratory angle $118^{o}$, $P_{lab}$
 = 400 GeV, for p--Li
and p--Ta collisions with
the String Fusion Model code results, with and without string fusion.

\newpage
\noindent{\Large {\bf Figure captions}}

\vskip 0.5cm

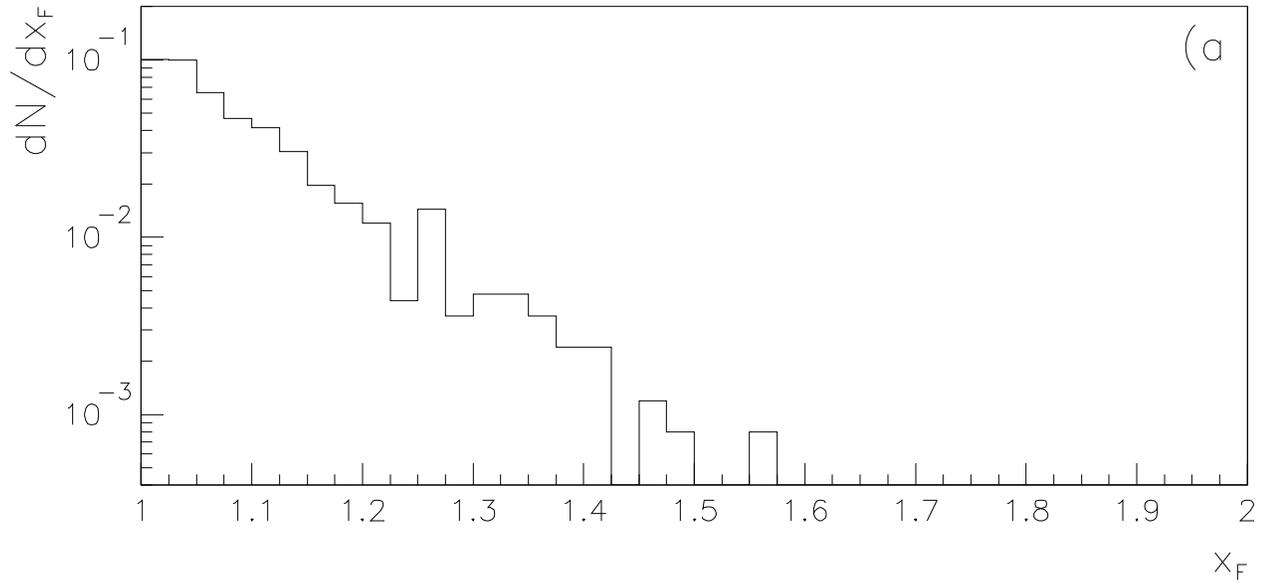
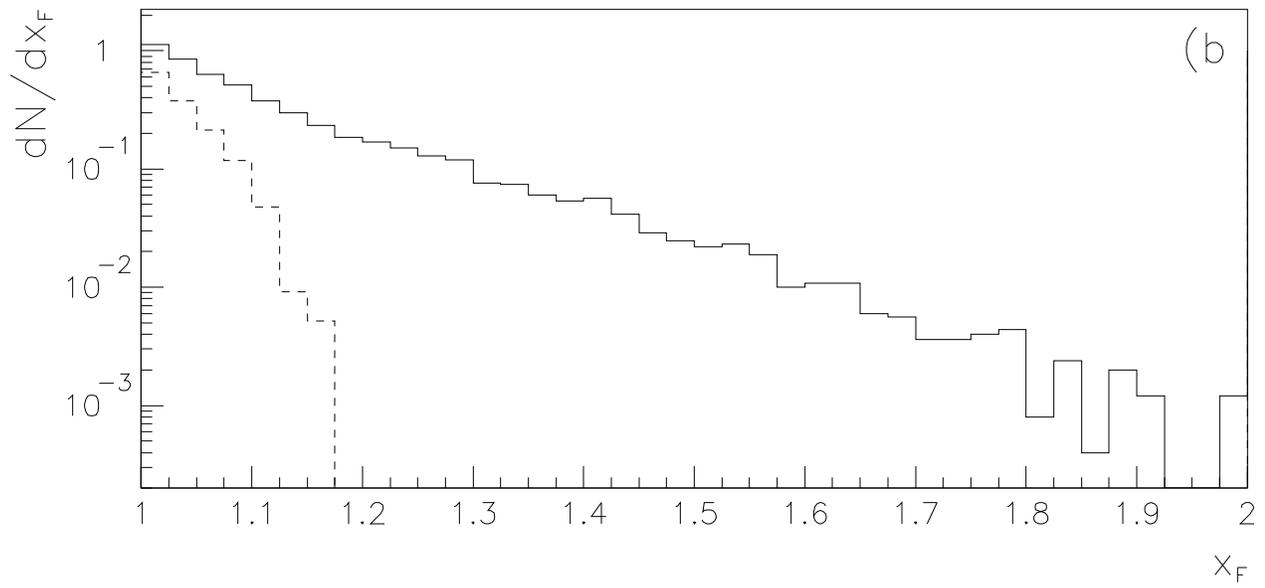
\noindent{\bf Fig. 1.} $x_F$ distributions for $x_F$ $>$ 1 in S--S 
collisions (10000 events) at $\sqrt{s}=19.4$ AGeV of mesons (a) and
baryons (b) with (continuous line) and without (dashed line)
string fusion. No mesons are found in the no fusion case.

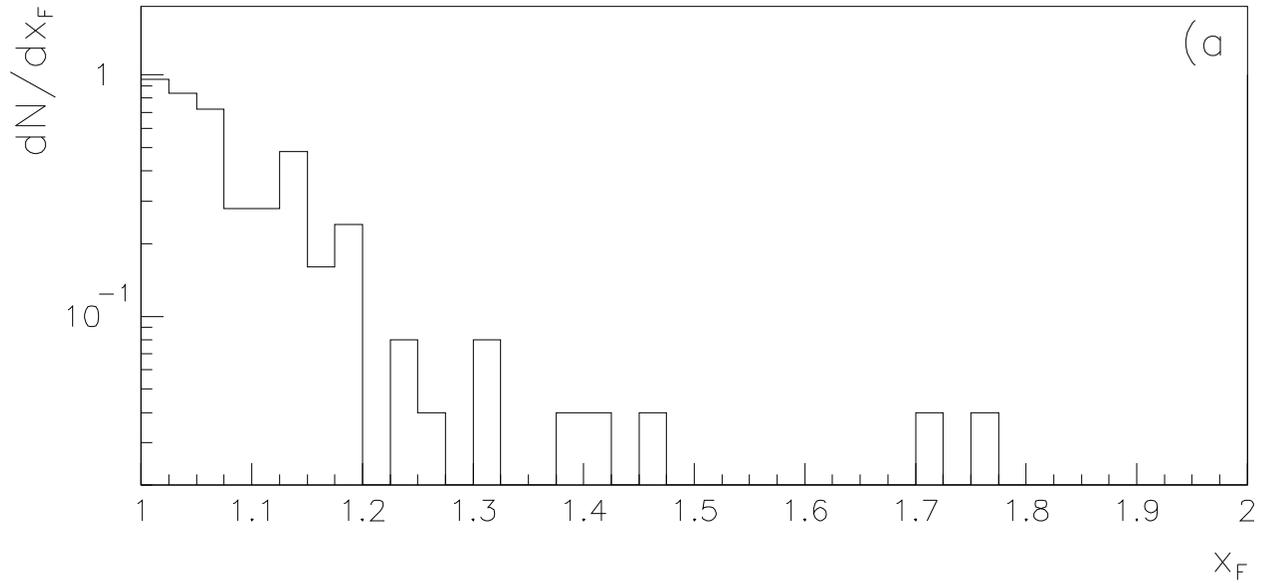
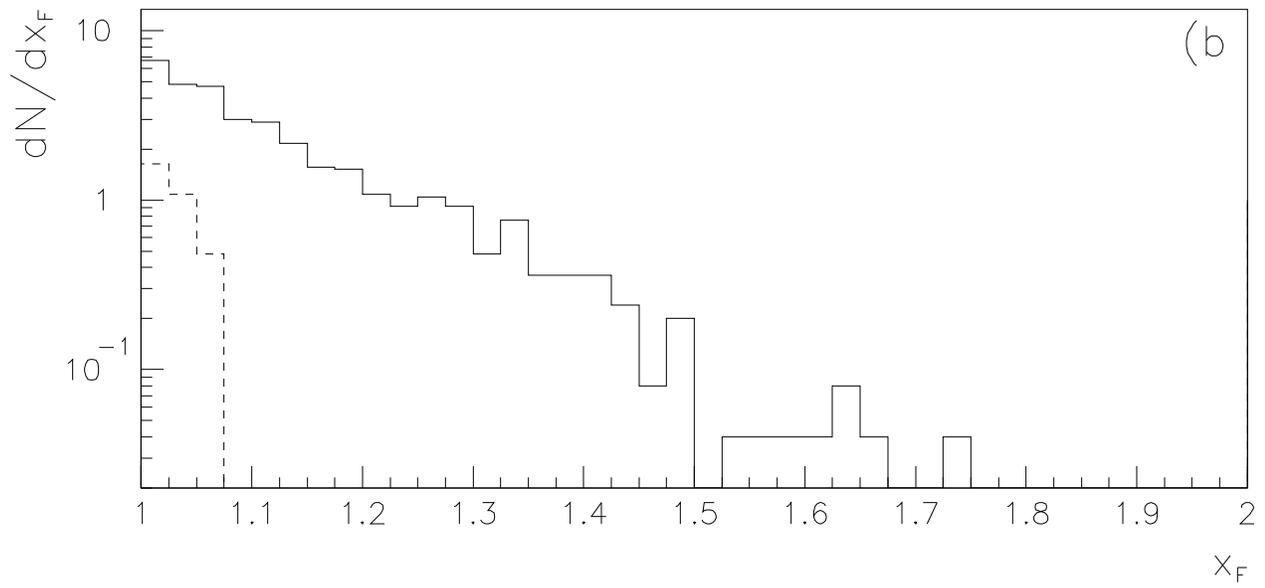
\noindent{\bf Fig. 2.} $x_F$ distributions for $x_F$ $>$ 1 in Pb--Pb 
collisions (1000 events) at $\sqrt{s}=19.4$ AGeV of mesons (a) and
baryons (b) with (continuous line) and without (dashed line)
string fusion. No mesons are found in the no fusion case.

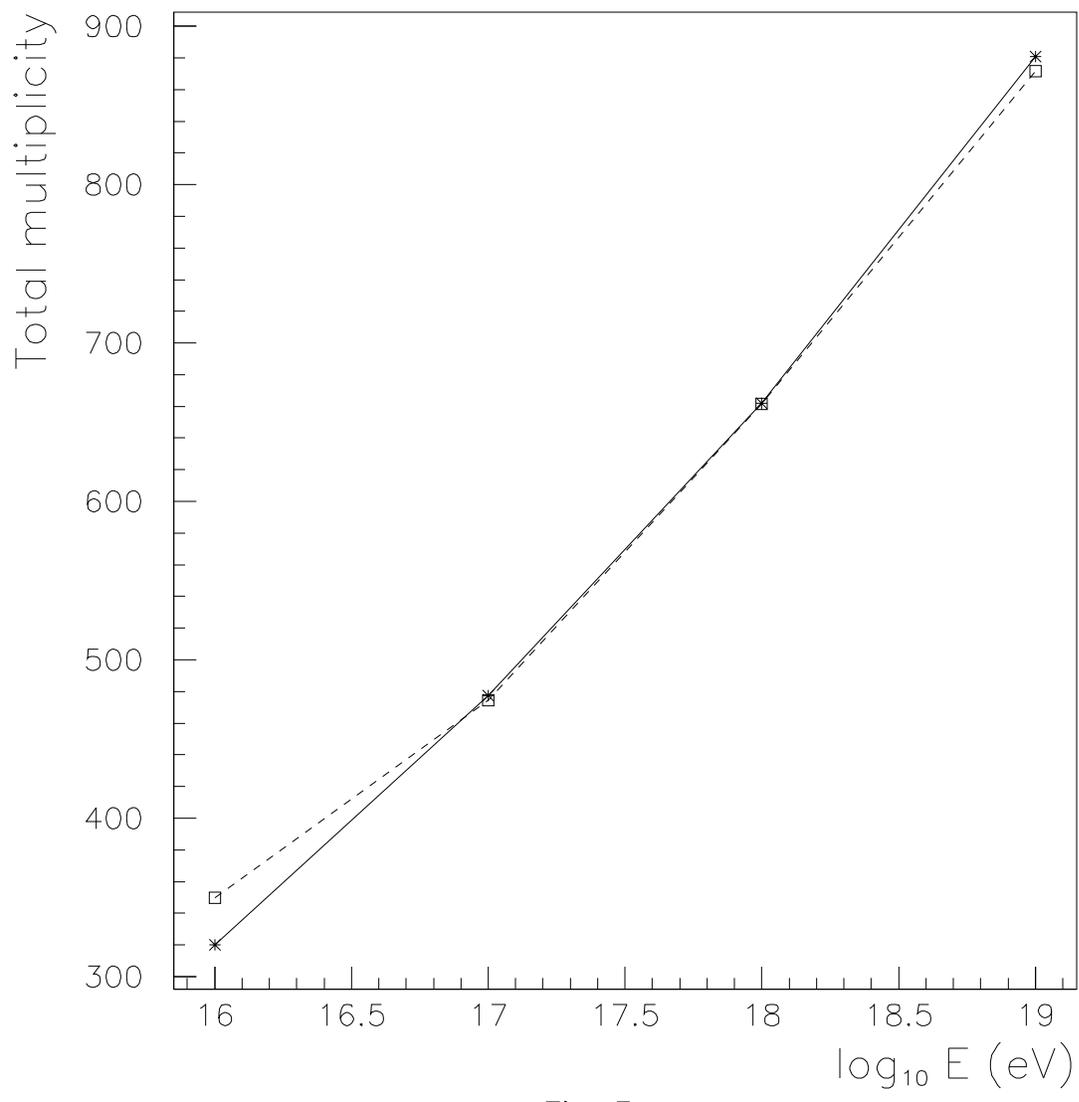
\noindent{\bf Fig. 3.} Total multiplicity dependence on the primary energy
for a fixed composition $<n_t>$ = 0.1 $<n_{p-Air}>$ + 0.9 $<n_{Fe-Air}>$
in the fusion case and a uniform change in the composition from 
$<n_t>$ = 0.25 $<n_{p-Air}>$ + 0.75 $<n_{Fe-Air}>$ at $10^{16}$ eV to
$<n_t>$ = 0.5 $<n_{p-Air}>$ + 0.5 $<n_{Fe-Air}>$ at $10^{19}$ eV
 in the no fusion case (dashed line).

\newpage

\begin{center}
{\bf Table 1}
\vskip 1cm
\begin{tabular}{cccc} \hline
\hline
Reaction & $p-Li$ & $p-Li$ & $p-Li$ \\ \hline
$p$ momentum & Experiment & Without fusion & With fusion  \\ \hline
& ${\sigma}$ for $\pi^{+}$ & ${\sigma}$ for $\pi^{+}$ 
& ${\sigma}$ for $\pi^{+}$ \\ \hline
0.200 & 5.75$\pm$0.79 & 7.06 & 9.6 \\ 
0.293 & 1.89$\pm$0.26 & 0.314 & 1.79 \\ 
0.381 & 0.672$\pm$0.046 & 0.11 & 0.53 \\ 
0.474 & 0.217$\pm$0.016 & 0 & 0.34 \\ 
0.580 & (0.509$\pm$0.044)$10^{-1}$ & 0 & 0.094 \\ 
0.681 & (0.128$\pm$0.012)$10^{-1}$ & 0 & 0.035 \\ \hline
\hline
Reaction & $p-Ta$ & $p-Ta$ & $p-Ta$ \\ \hline
$p$ momentum & Experiment & Without fusion & With fusion  \\ \hline
& ${\sigma}$ for $\pi^{+}$ & ${\sigma}$ for $\pi^{+}$
& ${\sigma}$ for $\pi^{+}$ \\ \hline
0.200 & 8.57$\pm$1.14 & 5.65 & 6.60 \\ 
0.293 & 2.20$\pm$0.31 & 0.19 & 1.57 \\ 
0.394 & 0.78$\pm$0.068 & 0.038 & 0.38 \\ 
0.489 & 0.309$\pm$0.032 & 0.032 & 0.173 \\ 
0.583 & 0.135$\pm$0.017 & 0 & 0.072 \\ 
0.680 & (0.386$\pm$0.076)$10^{-1}$ & 0 & 0.038 \\ \hline
& ${\sigma}$ for $K^{+}$ & ${\sigma}$ for $K^{+}$ 
& ${\sigma}$ for $K^{+}$ \\ \hline
0.539 & (0.241$\pm$0.100)$10^{-1}$ & 0 & 0.037 \\
0.584 & (0.372$\pm$0.763)$10^{-2}$ & 0 & 0 \\ \hline
\end{tabular}
\end{center}

\newpage

\begin{center}
{\bf Table 2}
\vskip 1cm
\begin{tabular}{cccc} \hline
\hline
Reaction & $p-Li$ & $p-Li$ & $p-Li$ \\ \hline
$p$ momentum & Experiment & Without fusion & With fusion  \\ \hline
& ${\sigma}$ for protons & ${\sigma}$ for protons
& ${\sigma}$ for protons \\ \hline
0.385 & 4.17$\pm$0.23 & 1.24 & 2.82 \\ 
0.476 & 1.76$\pm$0.11 & 0 & 1.01 \\ 
0.581 & 0.61$\pm$0.04 & 0 & 0.47 \\ \hline
\hline
Reaction & $p-Ta$ & $p-Ta$ & $p-Ta$ \\ \hline
$p$ momentum & Experiment & Without fusion & With fusion  \\ \hline
& ${\sigma}$ for protons & ${\sigma}$ for protons
& ${\sigma}$ for protons \\ \hline
0.395 & 29.9$\pm$1.5 & 15.1 & 22.3 \\ 
0.490 & 13.2$\pm$0.7 & 0 & 12.57 \\ 
0.585 & 5.2$\pm$0.3 & 0 & 2.5 \\ \hline
\end{tabular}
\end{center}


\begin{thebibliography}{99}

\bibitem{And87} B. Andersson, G. Gustafson and B. Nilsson--Almqvist, 
Nucl. Phys.
{\bf B281}, 289 (1987); M. Gyulassy, CERN preprint CERN--TH 4794 (1987). 

\bibitem{Cap94} A. Capella, U. P. Sukhatme, C.--I. Tan and J.
Tran Thanh Van, Phys. Rep. {\bf 236}, 225 (1994).

\bibitem{Wer93} K. Werner, Phys. Rep. {\bf 232}, 87 (1993).


\bibitem{Sor89} H. Sorge,
 H. St{\"o}cker and W. Greiner, Nucl. Phys. {\bf A498}, 567c (1989).

\bibitem{Paj90} C. Pajares, Proceedings of the International 
Workshop on Quark Gluon Signatures, Strasbourg, France,
 October 1--4 1990; M. A. Braun and C. Pajares, Phys. Lett. {\bf B287}, 154 
(1992);
 Nucl. Phys. {\bf B390}, 559 (1993); Nucl. Phys. {\bf B390}, 542 (1993).

\bibitem{And91} B. Andersson and P. Henning, Nucl. Phys. {\bf B355}, 
82 (1991).

\bibitem{Sor92} H. Sorge, M. Berenguer, H. St\"ocker and W. Greiner, Phys.
Lett. {\bf B289}, 6 (1992).

\bibitem{Moh92} H.--J. M\"{o}hring, J. Ranft, C. Merino and C. Pajares, Phys.
Rev. {\bf D47}, 4142 (1993); C. Merino, C. Pajares and J. Ranft, 
Phys. Lett. {\bf
B276}, 168 (1992).

\bibitem{Aic93} K. Werner and J. Aichelin, Phys. Lett. {\bf B308}, 372 (1993).

\bibitem{Ran94} J. Ranft, A. Capella and J. Tran Thanh Van, Phys. Lett. {\bf
B320}, 346 (1994).

\bibitem{Ame93}  N. S. Amelin, M. A. Braun and C. Pajares,
Phys. Lett. {\bf B306}, 312 (1993); Z. Phys. {\bf C63}, 507 (1994).

\bibitem{Arm94} 
N. Armesto, M. A. Braun, E. G. Ferreiro and C. Pajares, Phys. Lett. {\bf
B344}, 301 (1994).

\bibitem{Ele94} 
N. S. Amelin, N. Armesto, M. A. Braun, E. G. Ferreiro 
and C. Pajares, Phys. Rev. Lett. {\bf
73}, 2013 (1994).

\bibitem{Bir93} 
D. J. Bird {\it et al.}, Phys. Rev. Lett. {\bf 21}, 3401 (1993); Astrophys. J.
{\bf 42}, 491 (1994).

\bibitem{Gai93} T. K. Gaisser {\it et al.}, Phys. Rev. {\bf D47}, 1919 (1993).

\bibitem{Efi90} 
N. N. Efimov {\it et al.}, Proceedings of the  International 
Workshop on Astrophysical
Aspects of the Energetic Cosmic Rays, Kof\'u 1990.

\bibitem{Hay94} 
N. Hayshida  {\it et al.}, Phys. Rev. Lett. {\bf 73}, 3491 (1994); 
S. Yoshida {\it et al.}, Astropart. Phys. {\bf 3}, 105 (1995).

\bibitem{Gre66} 
K. Greisen, Phys. Rev. Lett. {\bf 16}, 748  
(1966); G. T. Zatsepin and V. A. Kuzmin, 
JETP Lett. {\bf 4}, 78 (1966).

\bibitem{Ste68} F. W. Stecker, Phys. Rev. Lett. {\bf 21}, 1016 (1968).

\bibitem{Sig11} G. Sigl, S. Lee, 
D. N. Schramm and P. Bhattacharjee, Univ. Chicago preprint (1996). 

\bibitem{Eng11} J. Engel, 
T. K. Gaisser, P. Lipari and T. Stanev, Phys. Rev. {\bf D46}, 5013 (1991).

\bibitem{Bal74} A. M. Baldin {\it et al.}, Yad. Fiz. {\bf 20}, 1210 (1974); 
Sov.
 J. Nucl. Phys. {\bf 20}, 629 (1975).

\bibitem{Str81} M. I. Strikman and L. L., Frankfurt Phys. Rep. {\bf 76}, 215  
(1981).

\bibitem{Efr88} A. V. Efremov, A. B. Kaidalov, V. T. Kim, G. I. Lykasov and 
N.
V. Slavin, Sov. J. Nucl. Phys. {\bf 47}, 868 (1988).

\bibitem{Bra94} M. A. Braun and V. Vechermin, Nucl. Phys. {\bf B427}, 614 
(1994).  

\bibitem{Bay79} Y. D. Bayukov {\it et al.}, Phys. Rev {\bf C20}, 764 (1979) 
, Phys. Rev. {\bf C22}, 700 (1980). 

\bibitem{Aug95} The Pierre Auger Proyect, Design Report, October 1995.

\end{thebibliography}
\end{document}